\documentclass[prd,aps, preprintnumbers, nofootinbib,superscriptaddress,notitlepage,twocolumn]{revtex4-2}
\usepackage[normalem]{ulem}
\usepackage{epsfig}
\usepackage{amsfonts}
\usepackage{amsmath}
\usepackage{slashed}
\usepackage{graphicx}
\usepackage{color}
\usepackage{mathtools}

\begin{document}
	
	\preprint{JLAB-THY-22-3569}
	
	\title{Inverse moment of the $B$-meson quasi distribution amplitude}

	\author{Ji Xu}
	\email{Corresponding author. xuji\_phy@zzu.edu.cn}
	\affiliation{School of Physics and Microelectronics, Zhengzhou University, Zhengzhou, Henan 450001, China}
	
	\author{Xi-Ruo Zhang}
	\email{Corresponding author. ZXRxiruo@163.com}
	\affiliation{School of Physics and Microelectronics, Zhengzhou University, Zhengzhou, Henan 450001, China}
	
	\author{Shuai Zhao}
	\email{Corresponding author. shzhao@jlab.org}
	\affiliation{Department of Physics, Old Dominion University, Norfolk, Virginia 23529, USA}
	\affiliation{Theory Center, Thomas Jefferson National Accelerator Facility, Newport News, Virginia 23606, USA}

	\begin{abstract}
		
		We perform a study on the structure of inverse moment (IM) of quasi distributions, by taking $B$-meson quasi distribution amplitude (quasi-DA) as an example. Based on a one-loop calculation, we derive the renormalization group equation and velocity evolution equation for the first IM of quasi-DA.
		We find that, in the large velocity limit, the first IM of $B$-meson quasi-DA can be factorized into IM as well as logarithmic moments of light-cone distribution amplitude (LCDA), accompanied by short distance coefficients.  Our results can be useful either in understanding the patterns of perturbative matching in Large Momentum Effective Theory or evaluating inverse moment of $B$-meson LCDA on the lattice.
		
	\end{abstract}
	
	\maketitle

	\section{Introduction}
	
	The structure of a high energy hadron can be depicted by nonperturbative functions like parton distribution functions (PDFs), light-cone distribution amplitudes (LCDA), etc. Because of their nonperturbative nature, parton distributions cannot be calculated with perturbation theory, instead, should be either extracted from experimental data or evaluated with nonperturbative methods like lattice QCD. However,  parton distributions are defined with matrix elements of nonlocal operators located on the light-cone, which cannot be simulated on the lattice.
	
	In recent years, it was pointed out that the difficulties of simulating parton physics on a Euclidean lattice can be overcome by employing the Large Momentum Effective Theory (LaMET) proposed by X. Ji~\cite{Ji:2013dva}. The idea underlying LaMET is to introduce quasidistributions, which are defined with matrix elements of equal-time nonlocal operators. The quasi distribution, when boosting to infinite momentum frame, i.e., $P_3\to \infty$ with $P_3$ being the hadron momentum on the moving direction, can be reduced to a light-cone distribution. The quasidistribution and its light-cone counterpart are related by a matching relation and the matching coefficient can be calculated with perturbative QCD since the difference between quasi and light-cone distributions is accompanied with a hard momentum scale $P_3\gg \Lambda_{\mathrm{QCD}}$.
	In recent years, LaMET has been applied on the lattice calculation of  PDFs, LCDAs, etc., for various hadrons. For recent reviews of LaMET, see e.g.~\cite{Cichy:2018mum,Ji:2020ect}. Other related approaches include pseudo distributions~\cite{Radyushkin:2017cyf,Radyushkin:2019mye}, lattice cross sections~\cite{Ma:2017pxb}, etc.
	
	It is interesting to study the moments of a quasi distribution.
	The positive moments, which are related to local operators, have been discussed in Refs.~\cite{Rossi:2017muf,Ji:2017rah,Radyushkin:2018nbf,Karpie:2018zaz}. For a quasi-PDF $\widetilde{f}(x,P_3)$, the asymptotic behavior at $|x|\to \infty$ is $~1/|x|$, thus the integral $\int dx x^n \widetilde{f}(x,P_3)$ leads to power divergence if $n$ is a non-negative integer; on the other hand, if $n$ is a negative number, the inverse moment (IM) defined by such integral has no power divergence.  Untill now, there are few studies on IM of quasidistributions, except the quasidistribution amplitude (quasi-DA) of heavy quarkonia~\cite{Jia:2015pxx}.
	
	In this work, we will study the IM of $B$-meson quasi-DA.
	The $B$-meson LCDA in LaMET is of particular interest. It is an inherent part of soft-collinear factorization theorems for many exclusive $B$ decay reactions~\cite{Beneke:2000ry,Beneke:2001ev,Beneke:2001at,DescotesGenon:2002mw,Bauer:2002aj,Beneke:2003zv,Becher:2005fg,Li:2012nk,Lu:2022fgz}; moreover, it is also an essential element in the light-cone sum-rule studies of the $B$-meson decays~~\cite{Khodjamirian:2006st,Faller:2008tr,Gubernari:2018wyi,Wang:2015vgv,Wang:2017jow,Lu:2018cfc,Gao:2019lta}. The $B$-meson LCDA has been studied in the framework of quasi distribution amplitude~\cite{Kawamura:2018gqz,Wang:2019msf} and the reduced Ioffe-time distribution~\cite{Zhao:2020bsx} approaches. However, there are still no lattice results.
It is still of great significance if the moment can be calculated from lattice QCD.	There are no positive moments for $B$-meson LCDA~\cite{Braun:2003wx}, but the IM exists. Moreover, the first IM is an indispensable part  of many factorization theorems in $B$ physics, e.g.,  the decay fraction of $B$-meson radiative decay~\cite{Beneke:2011nf,Wang:2016qii,Wang:2018wfj,Shen:2020hfq,Wang:2021yrr} and $B \to P, V$ form factors (see~\cite{Shen:2021yhe} for a recent review). Furthermore, the first IM is an essential parameter to build models for LCDA. The value of the first IM of $B$-meson LCDA has been estimated with various approaches, see, e.g., Refs.~\cite{Grozin:1996pq,Korchemsky:1999qb,Beneke:1999br,Ball:2003fq,Braun:2003wx}. However, there are few lattice-based calculations, and the precision has a large space to be improved.	
	It will be of phenomenological significance to study the IM of quasi-DA, which may shed light on the evaluation of LCDA on the lattice.

	In this Letter, we will introduce the IM of $B$-meson quasi DA and perform a theoretical study on the first IM of $B$-meson quasi-DA in LaMET. We will investigate the structure of IM of $B$-meson quasi-DA up to one-loop level. The renormalization, as well as one-loop matching between IMs of LCDA and quasi-DA, will be investigated. 
	
%

	\section{quasi distribution amplitude and inverse moment}\label{sec:def}
	
	We follow the notations in~\cite{Zhao:2020bsx}. To start with, let us consider a nonlocal heavy-light operator 	$	O_{\mu}(z,0;v)\equiv \bar q(z)\gamma_{\mu}\gamma_5 h_v(0)$ in heavy quark effective theory (HQET), 
	where $h_v$ is a heavy quark field in HQET, with $v$ denoting the velocity of $B$-meson. $v$ satisfies $v^2=1$ and $\slashed v h_v=h_v$; $\bar q (z)$ is a light quark field locating at $z$; $S(z,0)\equiv \operatorname{P}\exp [-i g\int_0^1 dtz_{\nu} A^{\nu}(tz)]$ is a Wilson line where $\operatorname{P}$ denotes the path ordering of operators. By analyzing the Lorentz structure of its meson-to-vacuum matrix element, we have
	\begin{align}
		&\left\langle 0\left|\bar q(z)S(z,0) \gamma_{\mu}\gamma_5  h_v(0)\right| \overline B( v)\right\rangle\nonumber\\
		=&i F(\mu)\left[v_{\mu} M_{B,v}(\nu, -z^2,\mu)+ z_{\mu} M_{B,z}(\nu, -z^2,\mu)\right], \label{eq:def}
	\end{align}
	where
	$M_{B,v}(\nu,\mu)$ and $M_{B,z}(\nu,\mu)$ are two scalar functions and $\nu\equiv v\cdot z$
	will be referred to
	as the ``Ioffe-time'' of the $B$-meson~\footnote{In QCD case, Ioffe-time is the inner product of momentum $p$ and $z$~\cite{Ioffe:1969kf,Braun:1994jq}.}.
	$F(\mu)$ is the decay constant of $B$-meson.
	$M_{B,v}$ term gives the twist-2 distribution when  $z^2\to 0$ while $M_{B,z}$ is a higher-twist contribution.
We
	rename the leading-twist function $M_{B,v}$ as $M_{B}$  and define $M_B(\nu,-z^2,\mu)$  as the Ioffe-time distribution amplitude (ITDA) of the $B$-meson for convenience.
	If $z$ is a light-like vector with minus component of $z$ being the only nonzero component, then  ITDA
	will reduce to the light-cone ITDA  $\mathcal{I}_B^+(\nu,\mu)$, i.e., $M_B(\nu,0,\mu)=\mathcal{I}_B^+(\nu,\mu)$, which is the LCDA in coordinate space.
	The $B$-meson LCDA is defined by the Fourier transform of $\mathcal{I}_B^+(\nu,\mu)$~\cite{Grozin:1996pq}.
	
	It was proposed in Refs.~\cite{Braun:2007wv,Ji:2013dva} that
	one can study equal-time separations
	$z=(0,0,0,z_3)$ on the lattice.
	The same idea has also been applied  for the  $B$-meson LCDA~\cite{Kawamura:2018gqz,Wang:2019msf,Zhao:2020bsx}. In this case, $\nu=-v_3 z_3$ and $z^2=-z_3^2$.
	The $B$-meson quasi-DA $\tilde{\phi}_B^+(\omega,v_3,\mu)$  can be expressed in terms of ITDA as~\cite{Kawamura:2018gqz,Wang:2019msf}
	\begin{align}
		\widetilde{\phi}_B^+ (\omega, v_3, \mu)=\frac{|v_3|}{2\pi}\int_{-\infty}^{\infty} dz_3 e^{i \omega v_3 z_3} M_B(-v_3 z_3,z_3^2,\mu)\, . \label{eq:qDA}
	\end{align}
	The matching relations linking the LCDA and quasi-DA or reduced-ITDA were derived in Refs.~\cite{Kawamura:2018gqz,Wang:2019msf,Zhao:2020bsx}.
	
	The first IM of LCDA is defined as
	\begin{align}
		\lambda_B^{-1}(\mu)\equiv \int_{-\infty}^{\infty}d\omega \frac{\phi_{B}^+(\omega,\mu)}{\omega}\, .\label{eq:IMLCDA}
	\end{align}
	Note that LCDA $\phi_{B}^+(\omega,\mu)$ only has nonzero support in $[0,\infty)$, hence the lower limit in Eq.~\eqref{eq:IMLCDA} is effectively $0$.
	Other important quantities in phenomenology are the logarithmic moments~\cite{Braun:2003wx}
	\begin{align}
		\sigma_n(\mu)=\lambda_B(\mu)\int_{-\infty}^{\infty}\frac{d\omega}{\omega}\ln^n \frac{\mu}{\omega} \phi_B^+(\omega,\mu)\, . \label{eq:log}
	\end{align}
	Similarly, we introduce IM of quasi-DA:
	\begin{align}
		\widetilde{\lambda}_B^{-1}(v_3,\mu)\equiv & \int_{-\infty}^{\infty}d\omega \frac{\widetilde{\phi}_{B}^+(\omega, v_3, \mu)}{\omega}\, . \label{eq:IMQDA}
	\end{align}
	In this case, however, the region of integration $(-\infty,\infty)$ is necessary because the support of quasi-DA extends to the whole axis.	
We note that the singularities at $\omega=0$ in the integrals Eqs.~\eqref{eq:IMLCDA},\eqref{eq:IMQDA}  need prescription. For future convenience, we add a small imaginary part $+i\epsilon$ to the denominator, i.e., $\omega \to \omega+i\epsilon$, to ensure that the IMs are well defined. However, IM of LCDA is independent of prescription because $\phi_B^+(\omega)\sim \omega$ when $\omega\to 0$~\cite{Lange:2003ff}.
	
	From Eqs.~\eqref{eq:def}, \eqref{eq:qDA} and \eqref{eq:IMQDA}, one can write down the operator definition of IM of $B$-meson quasi-DA as
\begin{align}
	\widetilde{\lambda}_B^{-1}&(v_3,\mu)|_{+i\epsilon}
	=	\frac{1}{ v_0 F(\mu)} \int_{0}^{-\infty} d \xi \nonumber\\
	&\cdot \left\langle 0\left|\bar q\left(\frac{\xi}{v_3}n\right)S\left(\frac{\xi}{v_3}n,0\right) \gamma_{0}\gamma_5  h_v(0)\right| \overline B( v)\right\rangle\, , \label{eq:IMopdef}
\end{align}	
where $n=(0,0,0,1)$ is the unit vector along the third direction. We choose the $0$-component of $\gamma_{\mu}$ so that the $M_{B,z}$ term is eliminated. ``$+i\epsilon$'' denotes that the above definition is based on the $+i\epsilon$ prescription. For $-i\epsilon$ prescription, the range of integral is from $0$ to $+\infty$. The definition under Cauchy's principal value is an average of $+i\epsilon$ and $-i\epsilon$ prescriptions. In the rest of this paper, the IM is under $+i\epsilon$ prescription  unless otherwise stated.  Because the inverse moment is defined with the matrix element of an equal-time operator, it can be simulated by lattice QCD directly, without calculating the whole quasi-DA first.
	
	

	\section{one-loop calculation}\label{sec:1loop}

	To study the renormalization and matching in perturbation theory, one can replace the hadron state with Fock state $|b\bar q\rangle$, as what we have done for DAs. Let $p$ being the momentum of the light quark, $p=\omega_0 v+p_{\perp}$ with $\omega_0>0$, where $v$ is the velocity of $B$-meson. We set $v=(v^0,0,0,v^3)$, or in light-cone coordinates, $v=(v^+, v^-,\mathbf{0}_{\perp})$. Assuming that the light quark is slightly offshell, i.e., $p^2<0$, then $-p^2$ will work as an infrared (IR) regulator. At tree level, we have $\phi_B^+(\omega,\omega_0)=\widetilde{\phi}_B^+(\omega,\omega_0)=\delta(\omega-\omega_0)$. This leads to the tree level result for IM: $1/\widetilde{\lambda}_B=1/\lambda_B=1/\omega_0$.

\begin{figure}
		\centering
		\includegraphics[width=0.8\linewidth]{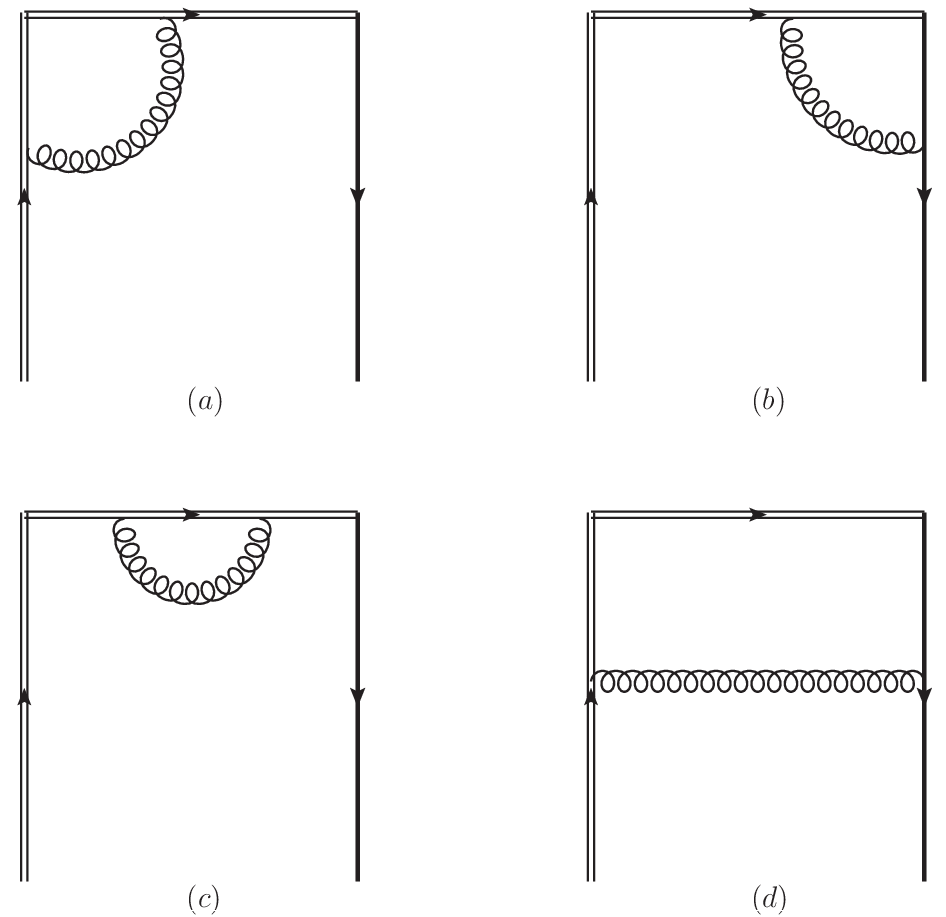}
		\caption{The Feynman diagrams for IMs of quasi-DA and LCDA.   The horizontal double line represents the gauge link, while the vertical double line denotes the heavy quark in HQET. The single line represents the light quark.}
		\label{fig:diagram}
	\end{figure}

	To evaluate the  one-loop corrections, we work in the Feynman gauge. The Feynman diagrams are shown in Fig.~\ref{fig:diagram}. Dimensional regularization (DR) is applied to regularize the UV singularities, where the space-time dimension is set to  $d=4-2\epsilon$.

	In fig.~\ref{fig:diagram}(a), there is a hard gluon exchange between the heavy quark and Wilson line.
	We have
	\begin{align}
		&\lambda_B^{-1}(\omega_0,\mu)|_{(a)}\nonumber\\
		=&	i  \frac{n\cdot v^2}{n\cdot p}\int\frac{d^d k}{(2\pi)^d}\frac{ g^2 \widetilde{\mu}^{2\epsilon} C_F  }{(k^2+i\epsilon) (-v\cdot k+i\epsilon) (n\cdot k+n\cdot p+i\epsilon)} \, ,
	\end{align}
	where $\widetilde\mu=\mu\sqrt{\frac{e^{\gamma_E}}{4\pi}}$, $\mu$ is an energy scale in DR and $\gamma_E$ is the Euler-Mascheroni constant; $C_F$ is a color factor with the value $4/3$.
	This integral can be easily calculated by using Feynman or Schwinger parameterizations. Since we have written the above expression in a Lorentz covariant form, it is applicable  for both light-cone and quasi cases.
	For IM of LCDA, we let $n$ being a light-cone vector and $n\cdot a=a^+$ for arbitrary vector $a$, $n^2=0$. We have
	\begin{align}
		&\lambda^{-1}_B(\omega_0, \mu)|_{(a)}\nonumber\\
		=&
		- \frac{\alpha_s C_F}{4\pi}  	\frac{1}{\omega_0 } \bigg(\frac{1}{\epsilon^2}+\frac{1}{\epsilon}\ln\frac{\mu^2}{\omega_0^2}+\frac{1}{2}\ln^2\frac{\mu^2}{\omega_0^2}+\frac{3\pi^2}{4}\bigg)\, .
	\end{align}
	On the other hand, for IM of quasi-DA, we choose $n$ as a unit vector in the 3rd direction of space-time,  $n\cdot a=a_3$ for an arbitrary vector $a$, and $n^2=-1$. After integration, one has
	\begin{align}
		\widetilde{\lambda}^{-1}_B&(\omega_0, v_3,\mu)|_{(a)}=
		-\frac{ \alpha_s C_F }{4\pi}\frac{1}{\omega_0  } \bigg[2\frac{\ln 2 i v_3}{ \epsilon}\nonumber\\
		&+2\ln 2i v_3 \bigg(\ln\frac{\mu^2}{ \omega_0^2}-\ln 2i v_3\bigg) +\frac{\pi^2}{3}\bigg]\, .
	\end{align}
	In the light-cone case one can observe a double pole $1/\epsilon^2$, which is the indication of cusp singularity; in the IM of quasi-DA, there is only single pole but accompanied with a meson-velocity-dependent coefficient. There are single and double logarithmic dependence on $v_3$ in the quasi case.

	Similar calculations can be performed to other diagrams.
	For fig.~\ref{fig:diagram}(b), one has, 
\begin{subequations}
	\begin{alignat}{2}
	\lambda_B^{-1}(\omega_0,\mu)|_{(b)}=&	  \frac{\alpha_s}{4\pi}C_F\frac{1}{\omega_0} \left(\frac{2}{\epsilon}+2\ln\frac{\mu^2}{-p^2}+4\right)\, ,\\
\widetilde{\lambda}_B^{-1}(\omega_0,\mu)|_{(b)}=&\frac{ \alpha_s }{4\pi}C_F \frac{1}{\omega_0}  \left(\frac{1}{\epsilon}+2\ln\frac{\mu^2}{-p^2}+2\ln 2i v_3\right.\nonumber\\
&~~~~~~\left.- \ln\frac{\mu^2}{\omega_0^2}+2 \right)\, .
\end{alignat}
\end{subequations}

	There is no contribution for light-cone case from fig.~\ref{fig:diagram}(c).
For IM of quasi-DA, the result reads
	\begin{align}
		&\widetilde{\lambda}_B^{-1}(\omega_0, v_3,\mu)|_{(c)}\nonumber\\
		= &    \frac{\alpha_s}{2\pi} C_F \frac{1}{\omega_0}    \left(-\frac{1}{\epsilon}+2\ln 2 i v_3 - \ln \frac{\mu^2}{\omega_0^2}-2\right)\, .
	\end{align}
	In the calculation of quasi-DA, Wilson line self-energy diagram Fig.~\ref{fig:diagram}(c) has a linear divergence.
	How to deal with the linear divergence is one of the major missions in the development of quasi and pseudo distribution approaches in the last few years. Many schemes, e.g., RI/MOM~\cite{Constantinou:2017sej,Stewart:2017tvs}, reduced Ioffe-time distribution~\cite{Orginos:2017kos,Radyushkin:2017lvu}, hybrid scheme~\cite{Ji:2020brr}, etc., have been proposed to renormalize the linear divergence in quasi distributions.  However, the IM
	involves only logarithmic UV divergence and has no linear divergence problem. It allows us to renormalize IM of quasi-DA with $\overline{\mathrm{MS}}$ scheme.

	For the box diagram fig.~\ref{fig:diagram}(d), we have
	\begin{align}
& i g^2 \widetilde{\mu}^{2\epsilon}  C_F \int\frac{d^d k}{(2\pi)^d}  \frac{1}{(k^2+i\epsilon) [(p+k)^2+i\epsilon] (-v\cdot k+i\epsilon)} \, .
	\end{align}
 This integral does not depend on $n$, indicating that it contributes the same results to IMs of LCDA and quasi-DA. Furthermore, the integral is UV finite. Therefore, the box diagram can be ignored both in renormalization and perturbative matching.
	In fact, it has already been shown in~\cite{Wang:2019msf,Zhao:2020bsx} that the box diagram does not contribute, both in the quasi and pseudo distribution approaches. Thus, it will not contribute to the renormalization and matching of IM as well. It can also be clarified through a $v_3$ power-counting argument.

	\section{Renormalization group equation}\label{sec:ren}
	
	We renormalize the IMs of LCDA and quasi-DA in $\overline{\mathrm{MS}}$ scheme:
\begin{subequations}
	\begin{alignat}{2}
		\lambda_{B,\mathrm{bare}}^{-1}(\omega_0)=&\bigg[1-\frac{\alpha_s(\mu)}{4\pi}C_F\bigg(\frac{1}{\epsilon^2}+\frac{1}{\epsilon}\ln\frac{\mu^2}{\omega_0^2}-\frac{2}{\epsilon}\bigg)\bigg]\nonumber\\
		&\lambda_{B}^{-1}(\omega_0;\mu)\, ,\\
		\widetilde{\lambda}_{B,\mathrm{bare}}^{-1}(\omega_0, v_3)=&\bigg[1-\frac{\alpha_s(\mu)}{4\pi}C_F \bigg(\frac{1}{\epsilon}+\frac{2}{\epsilon}\ln 2 i v_3\bigg)\bigg]\nonumber\\
		&\widetilde{\lambda}_{B}^{-1}(\omega_0;v_3,\mu)\, .
	\end{alignat}
\end{subequations}
	With the the above renormalization equation, we obtain RGE for IM of LCDA,
	\begin{align}
		\mu\frac{d}{d \mu}\lambda_B^{-1}(\mu)=
		&- \frac{\alpha_s(\mu)}{2\pi}C_F \left[2 \lambda_B^{-1}(\mu)\sigma_1(\mu)- \lambda_B^{-1}(\mu)\right]\, .
	\end{align}
	This result repeats the RGE derived in \cite{Ball:2003bf,Lange:2003ff}; it can also be derived from the Lange-Neubert equation for LCDA~\cite{Lange:2003ff}.
	On the  other hand, for IM of quasi-DA, one has
	\begin{align}
		\mu\frac{d}{d \mu}\widetilde{\lambda}_B^{-1}(v_3, \mu)=&- \frac{\alpha_s(\mu)}{2\pi}C_F(2\ln 2 i v_3+1)\widetilde{\lambda}_B^{-1}(v_3,\mu)\, .
	\end{align}
	One can observe that the IM of LCDA is not multiplicative renormalized, it will get mixed with logarithmic moment $\sigma_1$ at one-loop. However, for IM of quasi-DA, there is no mixing between 1st IM and logarithmic moments at one-loop level, which is different from IM of LCDA.  If one works at next-to-leading order accuracy, one can evolve $\widetilde{\lambda}_B$ to other scales without the input of other parameters.

	\section{Matching relation in LaMET and velocity RGE}

	In LaMET, light-cone and equal-time (``quasi'') quantities can be linked by a matching relation, with a perturbatively calculable hard function.	
	When $v_3\to \infty$, the equal-time matrix element that defines the IM of quasi-DA will become a light-cone matrix element, it indicates that under large Lorentz boost, $\widetilde{\lambda}_B(v_3,\mu)\to \lambda_B(\mu)$. With the spirit of LaMET,  one can expect the IR physics of $\widetilde{\lambda}_B(v_3,\mu)$ and $\lambda_B(\mu)$ are the same and there is a matching formula between IMs. Since the IMs have no dependence on $\omega$, a naive expectation for the matching relation is a multiplication instead of a convolution, i.e.,
	\begin{align}
		\widetilde{\lambda}_B(v_3,\mu_Q)=C\left(v_3, \mu_Q, \mu_L\right)\lambda_B(\mu_L)+\mathcal{O}(1/v_3)\, ,\label{eq:naive}
	\end{align}
	where $C(v_3, \mu_Q, \mu_L)$ is the hard coefficient, which can be expanded in series of $\alpha_s$ as $$C(v_3,\mu_Q, \mu_L)=\sum_{n=0}\left(\frac{\alpha_s(\mu)}{2\pi} C_F\right)^n  C^{(n)}(v_3,\mu_Q, \mu_L)\,;$$ $\mu_Q$ and $\mu_L$ are the scales that define the IMs of quasi and light-cone DAs, respectively.	
	We note that,
	a very similar multiplication-type matching relation was introduced in~\cite{Pilipp:2007sb}, which connects the IMs of LCDAs defined in HQET and full QCD; it can also be reproduced from the convolution-type matching between the LCDAs defined in QCD and HQET~\cite{Ishaq:2019dst,Zhao:2019elu}.	
	Another example is the matching of IMs of quasi-PDF and the normal PDF. One can start with the convolution-type matching relation for quasi PDF $\widetilde{f}$ and normal PDF $f$, then derive the matching relation for their IMs, which reads
\begin{align}
&	\int_{-\infty}^{\infty}\frac{dx}{x} \widetilde{f}(x, P_3, \mu_Q)\nonumber\\
		=& \int_{-\infty}^{\infty}\frac{dt}{ t} Z\left(t, P_3, \mu_Q, \mu_L\right)
		\cdot \int_{-1}^1 \frac{dy}{y} f(y, \mu_L)\, ,
\end{align}
which indicates that IM of quasi-PDF can be factorized as IM of normal PDF multiplied by a hard coefficient. 
	
	Expanding $\widetilde{\lambda}_B(\mu_Q)$, $\lambda_B(\mu_L)$ and $C(v_3, \mu_Q, \mu_L)$ in series of $\mathcal{O}(\alpha_s)$ in Eq.~\eqref{eq:naive}, up to one-loop level, we have $C^{(0)}=1$, and
	\begin{align}
		&C^{(1)}\left(v_3, \mu_Q, \mu_L\right)\nonumber\\
		=  &-\ln^2\frac{\mu_L}{\omega_0}+(2\ln 2i v_3+3) \ln\frac{\mu_L}{ \omega_0}+(2\ln 2i v_3+1) \ln\frac{\mu_Q}{\mu_L}\nonumber\\
		&-\ln 2 i v_3(\ln 2 i v_3+3) -\frac{5 \pi^2 }{24} +3  \, .
	\end{align}
	On the one hand, the IR singularities in IMs of LCDA and quasi-DA, which are regularized by $\ln(-p^2)$, are canceled, so the matching coefficient is IR free; on the other hand,
	$C^{(1)}(v_3,\mu_Q,\mu_L)$ depends on $\omega_0$. It means the matching coefficient depends on $\omega_0 v$---the momentum of external light quark, which indicates the failure of naive factorization relation Eq.~\eqref{eq:naive}; moreover, the single and double logarithmic dependence on $\omega_0$ should be absorbed into other nonperturbative quantities when $\omega_0$ is small. Thus the matching relation should not be a multiplication equation, which is different from the PDF case, and also the case of matching  IM in QCD to HQET~\cite{Pilipp:2007sb}. 
	
	Noticing that the $\ln^n \mu_L/\omega_0$ terms are related to the logarithmic moments defined in Eq.~\eqref{eq:log},
	we propose a modified factorization formula as
	\begin{align}
		\widetilde{\lambda}_B&(v_3, \mu_Q)=C_0\left(v_3,\frac{\mu_Q}{\mu_L}\right)\lambda_B(\mu_L)\nonumber\\
		&+\sum_{n=1}C_n \left(v_3,\frac{\mu_Q}{\mu_L}\right)\sigma_n (\mu_L)+\mathcal{O}(1/v_3)\, . \label{eq:modifac}
	\end{align}
	Again, the matching coefficients $C_n(v_3,\mu_Q/\mu_L)$ can be expanded in series of $\alpha_s$ as
$$
		C_n\left(v_3,\frac{\mu_Q}{\mu_L}\right)=\sum_{m=0}\left(\frac{\alpha_s(\mu)}{2\pi}C_F\right)^m C_n^{(m)}\left(v_3,\frac{\mu_Q}{\mu_L}\right)\, .
$$
	Performing the perturbative expansion, at the leading order, it gives $C_0^{(0)}=1$ and  $C_n^{(0)}=0$
	for $n \geq 1$.
	With one-loop results, we get the next-to-leading order corrections of the hard coefficients, which read
\begin{subequations}
	\begin{alignat}{4}
		C_0^{(1)}&\left(v_3,\frac{\mu_Q}{\mu_L}\right)=  (2\ln 2i v_3+1)\ln\frac{\mu_Q}{\mu_L}\nonumber\\
		&-\ln 2 i v_3 (\ln 2 i v_3+3)-\frac{5 \pi^2}{24}+3 \, ,\\
		C_1^{(1)}&\left(v_3,\frac{\mu_Q}{\mu_L}\right)=  2\ln 2 i v_3+3\, , \\
		C_2^{(1)}&\left(v_3,\frac{\mu_Q}{\mu_L}\right)= -1\, ,\\
		C_n^{(1)}&\left(v_3,\frac{\mu_Q}{\mu_L}\right)= 0\, ~~(n\geq 3)\ . 
	\end{alignat}\label{eq:matchcoe}
\end{subequations}

	{The matching relation Eqs.~\eqref{eq:modifac} and \eqref{eq:matchcoe} provide a convenient approach of extracting the inverse and logarithmic moments of LCDA. At one-loop order, because the matching coefficient $C_n^{(1)}$ are zero for $n\geq 3$, $\widetilde{\lambda}_B(v_3,\mu_Q)$ can be expressed as a second-order polynomial of $\ln 2i v_3$, with some coefficients involving the light-cone quantities $\lambda_B$, $\sigma_1$, $\sigma_2$. Thus one can calculate $\widetilde{\lambda}_B$ for several values of $v_3$ and then extract light-cone moments with a polynomial fit. Inversion of the matching formula is not necessary, and the only input is the first IM of quasi-DA.}

The matching relation indicates that the momentum evolution of $\widetilde{\lambda}_B(v_3,\mu)$ is only related to itself and the first logarithmic moment. In fact, one can write down
	the velocity evolution equation for $\widetilde{\lambda}_B(v_3, \mu)$ as
	\begin{align}
		v_3\frac{d}{d v_3}\widetilde{\lambda}_B(v_3,\mu)=&-\frac{\alpha_s}{2\pi}C_F\Big[(2\ln 2i v_3+3)\widetilde{\lambda}_B(v_3,\mu)\nonumber\\
		&-2\widetilde{\sigma}_1(v_3, \mu)\Big]\, .
	\end{align}
	To evolve $\widetilde{\lambda}_B(v_3,\mu)$ from one velocity to another, one needs the input of the first IM and logarithmic moment.
	
	At last, one may notice that the matching coefficients $C_{n}$ are complex numbers, therefore, the IM of quasi-DA is also complex. This is due to the $+i\epsilon$ prescription we employed at $\omega=0$. One can also get  real values for IM and matching coefficients with  other  prescriptions, e.g, Cauchy principal value. Since  $\operatorname{P.V.}(1/\omega)=\frac12(\frac{1}{\omega+i\epsilon}+\frac{1}{\omega-i\epsilon})$, one can get $\widetilde\lambda_B$ under principal value prescription by averaging the results under $+i\epsilon$ prescription and their complex conjugates, and this is equivalent to taking the real parts in our results.  It is verified by a direct calculation under principal value prescription.

	\section{discussions and summary}\label{sec:sum}

	We introduce the inverse moment of $B$-meson quasi-DA and explore its properties.  	The IM of quasi-DA can be simulated on a Euclidean lattice since it is defined with an integral of equal-time matrix elements. 
	With a one-loop calculation of IMs, we derive the RGE for the first IM of quasi-DA and figure out the correct form of matching relation in LaMET. The first IM of quasi-DA is not only factorized into the first IM  but also the logarithmic moments of LCDA, accompanied by hard coefficients which are free of IR singularities and independent of external states.

	Distinguished from non-negative moments,  IM is not defined by a local operator, so the determination of IM needs the  matrix element for all $z$.  However, our  approach in this work
	 has some advantages in practical calculations. According to the definition of IM in Eq.~\eqref{eq:IMopdef}, IM of quasi-DA can be calculated with an integral of the equal-time matrix element, which can be approximated by a summation over discrete points when $\xi$ is small; on the other hand, one can extrapolate the large $\xi$ contribution with the help of long-distance asymptotic behavior of spatial correlation function, and perform the integration in this region analytically~\cite{Ji:2020brr}. Then, because the matching relation is a linear combination instead of a convolution, the errors and unphysical oscillations due to Fourier transform and convolution can be avoided. The mixing of IM and logarithmic moments of LCDA in the matching formula may cause other difficulties.
	 Such difficulties caused by mixing may be overcome by evaluating $\widetilde{\lambda}_B$ with several $v_3$ and extracting the IM and logarithmic moments by fitting the $v_3$ dependence.

Furthermore, because the lattice simulation of nonlocal HQET matrix element is very challenging, a realistic idea may be building proper models for quasi-DA and fitting the parameters of models with a few lattice data.  The IM of quasi-DA will be an essential parameter for quasi-DA models, just like the IM for the models of LCDA ~\cite{Grozin:1996pq,Braun:2003wx}. Further lattice simulations on IM of quasi-DA will be crucial to improve the determination of IM and other parameters of $B$-meson LCDA.

	\section*{Acknowledgments}
	
    We thank Wei Wang for careful reading of the manuscript and suggestions, and Yu-Ming Wang for valuable comments. The work of J.~X. and X.~R.~Z. is supported in part by the National Natural Science Foundation of China under Grant No.~12105247 and No.~12047545, the China Postdoctoral Science Foundation under Grant No.~2021M702957. The work  of S.~Z. is supported by Jefferson Science Associates,
	LLC under  U.S. DOE Contract \# DE-AC05-06OR23177
	and by U.S. DOE Grant \# DE-FG02-97ER41028.
	
	%
	%
	%
	%


\begin{thebibliography}{widestlabel}
		
\bibitem{Ji:2013dva}
X.~Ji,
Phys. Rev. Lett. \textbf{110}, 262002 (2013)
doi:10.1103/PhysRevLett.110.262002
[arXiv:1305.1539 [hep-ph]].

\bibitem{Cichy:2018mum}
K.~Cichy and M.~Constantinou,
Adv. High Energy Phys. \textbf{2019}, 3036904 (2019)
doi:10.1155/2019/3036904
[arXiv:1811.07248 [hep-lat]].

\bibitem{Ji:2020ect}
X.~Ji, Y.~Liu, Y.~S.~Liu, J.~H.~Zhang and Y.~Zhao,
Rev. Mod. Phys. \textbf{93}, no.3, 035005 (2021)
doi:10.1103/RevModPhys.93.035005
[arXiv:2004.03543 [hep-ph]].

\bibitem{Radyushkin:2017cyf}
A.~V.~Radyushkin,
Phys. Rev. D \textbf{96}, no.3, 034025 (2017)
doi:10.1103/PhysRevD.96.034025
[arXiv:1705.01488 [hep-ph]].

\bibitem{Radyushkin:2019mye}
A.~V.~Radyushkin,
Int. J. Mod. Phys. A \textbf{35}, no.05, 2030002 (2020)
doi:10.1142/S0217751X20300021
[arXiv:1912.04244 [hep-ph]].

\bibitem{Ma:2017pxb}
Y.~Q.~Ma and J.~W.~Qiu,
Phys. Rev. Lett. \textbf{120}, no.2, 022003 (2018)
doi:10.1103/PhysRevLett.120.022003
[arXiv:1709.03018 [hep-ph]].

\bibitem{Rossi:2017muf}
G.~C.~Rossi and M.~Testa,
Phys. Rev. D \textbf{96}, no.1, 014507 (2017)
doi:10.1103/PhysRevD.96.014507
[arXiv:1706.04428 [hep-lat]].

\bibitem{Ji:2017rah}
X.~Ji, J.~H.~Zhang and Y.~Zhao,
Nucl. Phys. B \textbf{924}, 366-376 (2017)
doi:10.1016/j.nuclphysb.2017.09.001
[arXiv:1706.07416 [hep-ph]].

\bibitem{Radyushkin:2018nbf}
A.~V.~Radyushkin,
Phys. Lett. B \textbf{788}, 380-387 (2019)
doi:10.1016/j.physletb.2018.11.047
[arXiv:1807.07509 [hep-ph]].

\bibitem{Karpie:2018zaz}
J.~Karpie, K.~Orginos and S.~Zafeiropoulos,
JHEP \textbf{11}, 178 (2018)
doi:10.1007/JHEP11(2018)178
[arXiv:1807.10933 [hep-lat]].

\bibitem{Jia:2015pxx}
Y.~Jia and X.~Xiong,
Phys. Rev. D \textbf{94}, no.9, 094005 (2016)
doi:10.1103/PhysRevD.94.094005
[arXiv:1511.04430 [hep-ph]].

\bibitem{Beneke:2000ry}
M.~Beneke, G.~Buchalla, M.~Neubert and C.~T.~Sachrajda,
Nucl. Phys. B \textbf{591}, 313-418 (2000)
doi:10.1016/S0550-3213(00)00559-9
[arXiv:hep-ph/0006124 [hep-ph]].

\bibitem{Beneke:2001ev}
M.~Beneke, G.~Buchalla, M.~Neubert and C.~T.~Sachrajda,
Nucl. Phys. B \textbf{606}, 245-321 (2001)
doi:10.1016/S0550-3213(01)00251-6
[arXiv:hep-ph/0104110 [hep-ph]].

\bibitem{Beneke:2001at}
M.~Beneke, T.~Feldmann and D.~Seidel,
Nucl. Phys. B \textbf{612}, 25-58 (2001)
doi:10.1016/S0550-3213(01)00366-2
[arXiv:hep-ph/0106067 [hep-ph]].

\bibitem{DescotesGenon:2002mw}
S.~Descotes-Genon and C.~T.~Sachrajda,
Nucl. Phys. B \textbf{650}, 356-390 (2003)
doi:10.1016/S0550-3213(02)01066-0
[arXiv:hep-ph/0209216 [hep-ph]].

\bibitem{Bauer:2002aj}
C.~W.~Bauer, D.~Pirjol and I.~W.~Stewart,
Phys. Rev. D \textbf{67}, 071502(R) (2003)
doi:10.1103/PhysRevD.67.071502
[arXiv:hep-ph/0211069 [hep-ph]].

\bibitem{Beneke:2003zv}
M.~Beneke and M.~Neubert,
Nucl. Phys. B \textbf{675}, 333-415 (2003)
doi:10.1016/j.nuclphysb.2003.09.026
[arXiv:hep-ph/0308039 [hep-ph]].

\bibitem{Becher:2005fg}
T.~Becher, R.~J.~Hill and M.~Neubert,
Phys. Rev. D \textbf{72}, 094017 (2005)
doi:10.1103/PhysRevD.72.094017
[arXiv:hep-ph/0503263 [hep-ph]].

\bibitem{Li:2012nk}
H.~n.~Li, Y.~L.~Shen and Y.~M.~Wang,
Phys. Rev. D \textbf{85}, 074004 (2012)
doi:10.1103/PhysRevD.85.074004
[arXiv:1201.5066 [hep-ph]].

\bibitem{Lu:2022fgz}
C.~D.~L\"u, Y.~L.~Shen, C.~Wang and Y.~M.~Wang,
[arXiv:2202.08073 [hep-ph]].

\bibitem{Khodjamirian:2006st}
A.~Khodjamirian, T.~Mannel and N.~Offen,
Phys. Rev. D \textbf{75}, 054013 (2007)
doi:10.1103/PhysRevD.75.054013
[arXiv:hep-ph/0611193 [hep-ph]].

\bibitem{Faller:2008tr}
S.~Faller, A.~Khodjamirian, C.~Klein and T.~Mannel,
Eur. Phys. J. C \textbf{60}, 603-615 (2009)
doi:10.1140/epjc/s10052-009-0968-4
[arXiv:0809.0222 [hep-ph]].

\bibitem{Gubernari:2018wyi}
N.~Gubernari, A.~Kokulu and D.~van Dyk,
JHEP \textbf{01}, 150 (2019)
doi:10.1007/JHEP01(2019)150
[arXiv:1811.00983 [hep-ph]].

\bibitem{Wang:2015vgv}
Y.~M.~Wang and Y.~L.~Shen,
Nucl. Phys. B \textbf{898}, 563-604 (2015)
doi:10.1016/j.nuclphysb.2015.07.016
[arXiv:1506.00667 [hep-ph]].

\bibitem{Wang:2017jow}
Y.~M.~Wang, Y.~B.~Wei, Y.~L.~Shen and C.~D.~L\"u,
JHEP \textbf{06}, 062 (2017)
doi:10.1007/JHEP06(2017)062
[arXiv:1701.06810 [hep-ph]].

\bibitem{Lu:2018cfc}
C.~D.~L\"u, Y.~L.~Shen, Y.~M.~Wang and Y.~B.~Wei,
JHEP \textbf{01}, 024 (2019)
doi:10.1007/JHEP01(2019)024
[arXiv:1810.00819 [hep-ph]].

\bibitem{Gao:2019lta}
J.~Gao, C.~D.~L\"u, Y.~L.~Shen, Y.~M.~Wang and Y.~B.~Wei,
Phys. Rev. D \textbf{101}, no.7, 074035 (2020)
doi:10.1103/PhysRevD.101.074035
[arXiv:1907.11092 [hep-ph]].

\bibitem{Kawamura:2018gqz}
H.~Kawamura and K.~Tanaka,
PoS \textbf{RADCOR2017}, 076 (2018)
doi:10.22323/1.290.0076

\bibitem{Wang:2019msf}
W.~Wang, Y.~M.~Wang, J.~Xu and S.~Zhao,
Phys. Rev. D \textbf{102}, no.1, 011502(R) (2020)
doi:10.1103/PhysRevD.102.011502
[arXiv:1908.09933 [hep-ph]].

\bibitem{Zhao:2020bsx}
S.~Zhao and A.~V.~Radyushkin,
Phys. Rev. D \textbf{103}, no.5, 054022 (2021)
doi:10.1103/PhysRevD.103.054022
[arXiv:2006.05663 [hep-ph]].

\bibitem{Braun:2003wx}
V.~M.~Braun, D.~Y.~Ivanov and G.~P.~Korchemsky,
Phys. Rev. D \textbf{69}, 034014 (2004)
doi:10.1103/PhysRevD.69.034014
[arXiv:hep-ph/0309330 [hep-ph]].

\bibitem{Beneke:2011nf}
M.~Beneke and J.~Rohrwild,
Eur. Phys. J. C \textbf{71}, 1818 (2011)
doi:10.1140/epjc/s10052-011-1818-8
[arXiv:1110.3228 [hep-ph]].

\bibitem{Wang:2016qii}
Y.~M.~Wang,
JHEP \textbf{09}, 159 (2016)
doi:10.1007/JHEP09(2016)159
[arXiv:1606.03080 [hep-ph]].

\bibitem{Wang:2018wfj}
Y.~M.~Wang and Y.~L.~Shen,
JHEP \textbf{05}, 184 (2018)
doi:10.1007/JHEP05(2018)184
[arXiv:1803.06667 [hep-ph]].

\bibitem{Shen:2020hfq}
Y.~L.~Shen, Y.~M.~Wang and Y.~B.~Wei,
JHEP \textbf{12}, 169 (2020)
doi:10.1007/JHEP12(2020)169
[arXiv:2009.02723 [hep-ph]].

\bibitem{Wang:2021yrr}
C.~Wang, Y.~M.~Wang and Y.~B.~Wei,
JHEP \textbf{02}, 141 (2022)
doi:10.1007/JHEP02(2022)141
[arXiv:2111.11811 [hep-ph]].

\bibitem{Shen:2021yhe}
Y.~L.~Shen and Y.~B.~Wei,
Adv. High Energy Phys. \textbf{2022}, 2755821 (2022)
doi:10.1155/2022/2755821
[arXiv:2112.01500 [hep-ph]].

\bibitem{Grozin:1996pq}
A.~G.~Grozin and M.~Neubert,
Phys. Rev. D \textbf{55}, 272-290 (1997)
doi:10.1103/PhysRevD.55.272
[arXiv:hep-ph/9607366 [hep-ph]].

\bibitem{Korchemsky:1999qb}
G.~P.~Korchemsky, D.~Pirjol and T.~M.~Yan,
Phys. Rev. D \textbf{61}, 114510 (2000)
doi:10.1103/PhysRevD.61.114510
[arXiv:hep-ph/9911427 [hep-ph]].

\bibitem{Beneke:1999br}
M.~Beneke, G.~Buchalla, M.~Neubert and C.~T.~Sachrajda,
Phys. Rev. Lett. \textbf{83}, 1914-1917 (1999)
doi:10.1103/PhysRevLett.83.1914
[arXiv:hep-ph/9905312 [hep-ph]].

\bibitem{Ball:2003fq}
P.~Ball and E.~Kou,
JHEP \textbf{04}, 029 (2003)
doi:10.1088/1126-6708/2003/04/029
[arXiv:hep-ph/0301135 [hep-ph]].

\bibitem{Ioffe:1969kf}
B.~L.~Ioffe,
Phys. Lett. B \textbf{30}, 123-125 (1969)
doi:10.1016/0370-2693(69)90415-8

\bibitem{Braun:1994jq}
V.~Braun, P.~Gornicki and L.~Mankiewicz,
Phys. Rev. D \textbf{51}, 6036-6051 (1995)
doi:10.1103/PhysRevD.51.6036
[arXiv:hep-ph/9410318 [hep-ph]].

\bibitem{Braun:2007wv}
V.~Braun and D.~M\"uller,
Eur. Phys. J. C \textbf{55}, 349-361 (2008)
doi:10.1140/epjc/s10052-008-0608-4
[arXiv:0709.1348 [hep-ph]].

\bibitem{Lange:2003ff}
B.~O.~Lange and M.~Neubert,
Phys. Rev. Lett. \textbf{91}, 102001 (2003)
doi:10.1103/PhysRevLett.91.102001
[arXiv:hep-ph/0303082 [hep-ph]].

\bibitem{Constantinou:2017sej}
M.~Constantinou and H.~Panagopoulos,
Phys. Rev. D \textbf{96}, no.5, 054506 (2017)
doi:10.1103/PhysRevD.96.054506
[arXiv:1705.11193 [hep-lat]].

\bibitem{Stewart:2017tvs}
I.~W.~Stewart and Y.~Zhao,
Phys. Rev. D \textbf{97}, no.5, 054512 (2018)
doi:10.1103/PhysRevD.97.054512
[arXiv:1709.04933 [hep-ph]].

\bibitem{Orginos:2017kos}
K.~Orginos, A.~Radyushkin, J.~Karpie and S.~Zafeiropoulos,
Phys. Rev. D \textbf{96}, no.9, 094503 (2017)
doi:10.1103/PhysRevD.96.094503
[arXiv:1706.05373 [hep-ph]].

\bibitem{Radyushkin:2017lvu}
A.~V.~Radyushkin,
Phys. Lett. B \textbf{781}, 433-442 (2018)
doi:10.1016/j.physletb.2018.04.023
[arXiv:1710.08813 [hep-ph]].

\bibitem{Ji:2020brr}
X.~Ji, Y.~Liu, A.~Sch\"afer, W.~Wang, Y.~B.~Yang, J.~H.~Zhang and Y.~Zhao,
Nucl. Phys. B \textbf{964}, 115311 (2021)
doi:10.1016/j.nuclphysb.2021.115311
[arXiv:2008.03886 [hep-ph]].

\bibitem{Ball:2003bf}
P.~Ball,
[arXiv:hep-ph/0308249 [hep-ph]].

\bibitem{Pilipp:2007sb}
V.~Pilipp,
[arXiv:hep-ph/0703180 [hep-ph]].

\bibitem{Ishaq:2019dst}
S.~Ishaq, Y.~Jia, X.~Xiong and D.~S.~Yang,
Phys. Rev. Lett. \textbf{125}, no.13, 132001 (2020)
doi:10.1103/PhysRevLett.125.132001
[arXiv:1905.06930 [hep-ph]].

\bibitem{Zhao:2019elu}
S.~Zhao,
Phys. Rev. D \textbf{101}, no.7, 071503(R) (2020)
doi:10.1103/PhysRevD.101.071503
[arXiv:1910.03470 [hep-ph]].
		
	\end{thebibliography}
\end{document}